_______________________________________________________________

# Multivariate information measures:  a unification using Möbius operators on subset lattices
_______________________________________________________________

David J. Galas[1*]  and Nikita A. Sakhanenko[1]

[1]Pacific Northwest Research Institute
720 Broadway
Seattle, Washington 98122
USA

\* correspondence to:   David Galas, dgalas@pnri.org

# **Abstract**

Information-related measures are useful tools for multi-variable data analysis, as measures of dependence among variables, and as descriptions of order in biological and physical systems. Information-related measures, like marginal entropies, mutual / interaction / multi -information, have been used in a number of fields including descriptions of systems complexity and biological data analysis. The mathematical relationships among these measures are therefore of significant interest. Relations between common information measures include the duality relations based on Möbius inversion on lattices. These are the direct consequence of the symmetries of the lattices of the sets of variables (subsets ordered by inclusion). While the mathematical properties and relationships among these information-related measures are of significant interest, there has been, to our knowledge, no systematic examination of the full range of relationships and no unification of this diverse range of functions into a single formalism as we do here. In this paper we define operators on functions on these lattices based on the Möbius inversion idea that map the functions into one another (Möbius operators.) We show that these operators form a simple group isomorphic to the symmetric group $S_3$. Relations among the set of functions on the lattice are transparently expressed in terms of the operator algebra, and, applied to the information measures, can be used to derive a wide range of relationships among measures. We describe a direct relation between sums of conditional log-likelihoods and previously defined dependency measures. The algebra is naturally generalized which yields more extensive relationships. This formalism provides a fundamental unification of information-related measures, but isomorphism of all distributive lattices with the subset lattice implies broad potential application of these results.

**Introduction**

The description of order and disorder in systems of all kinds is fundamental. In the physics and chemistry of condensed matter it plays a central role, but for systems with biological levels of complexity, including interactions of genes, macromolecules, cells and of networks of neurons, it is also central, and certainly not well understood. Mathematical descriptions of the underlying order, and transitions between states of order, are still far from satisfactory and a subject of much current research. The difficulty arises in several ways, but the dominant contributors are probably the high degree of effective interactions and their non-linearity. There have been many efforts to define information-based measures as a language for describing the order and disorder of systems and the transfer of information. Negative entropy, joint entropies, multi-information and various manifestations of Kullback-Leibler divergence are among the key concepts. Interaction information is one of these. It is an entropy-based measure for multiple variables introduced by McGill in 1954 [1]. It has been used effectively in a number of developments and applications of information-based analysis [2-5], and has several interesting properties, including symmetry under permutation of variables, like joint entropies and multi-information, though its interpretation as a form of information in the usual sense is ambiguous as it can have negative values. In previous work we have proposed complexity and dependence measures related to this quantity [6,9]. Here we focus on elucidating the character and source of some of the mathematical properties that relate these measures, and on extending both the definitions and spectrum of relations among all these quantities. The formalism presented here can thus be viewed as a unification of a wide range of information-related measures in the sense that the relations between them are elucidated.

At the two variable level multi-information, K-L divergence and interaction information are all identical, and equal to mutual information. The interaction information $I(v_n)$ for a set of $n$ variables or attributes, $v_n = \{X_1, X_2, X_3... X_n\}$, obeys a recursion relation that parallels that for the joint entropy of sets of variables, $H(v_n)$, which is derived in turn directly from the probability chain rule:

$$H(v_n) = H(v_{n-1}) + H(X_n \mid v_{n-1})$$

$$\tag{1}$$

$$I(v_n) = I(v_{n-1}) - I(v_{n-1} \mid X_n)$$

where the second terms on the right are conditionals. These two information functions are known to be related by Möbius inversion [2-5]. There is an inherent duality between the marginal entropy functions and the interaction information functions based on Möbius inversion. Bell described an elegantly symmetric form of the inversion, and identified the source of this duality in the lattice associated with the variables [2]. The duality is based on the partially ordered set of variables, subsets ordered by inclusion, which corresponds to its power set lattice. We start with this symmetric inversion relation and extend it to an algebra of operators on these lattices.

This paper is structured as follows. We briefly review the definitions relevant to Möbius inversion, and define the operators that map the functions on the lattice into one another, expressing the Möbius inversions as operator equations. We then determine the products of the operators and, completing the set of operators with a lattice complement operator, we show that they form a group that is isomorphic to the symmetric group, $S_3$. In the next section we express previous results in defining dependency and complexity measures in terms of the operator formalism, and illustrate relationships between many commonly used information measures, like total correlation or multi-information. We derive a number of new relations using the formalism, and point out the relationship between multi-information and certain maximum entropy limits. This suggests a wide range of maximum entropy criteria in the relationships inherent in the operator algebra.

The next section focuses on the relations between these functions and the probability distributions underlying the symmetries. We then illustrate an operator equation relating our dependence measure to conditional log likelihood functions. Finally, we define a generalized form of the inversion relation, which also has $S_3$ symmetry, and show how these operators on functions can be additively decomposed in a variety of ways.

### 1. Möbius Dualities

Many applications make use of the relations among information theoretic quantities like joint entropies and interaction information that are formed by what can be called Möbius duality [2]. Restricting ourselves to functions on subset lattices, we note that a function on a subset lattice is a mapping of each of the elements subsets to the reals. The Möbius function for this poset ordered by inclusion is $\mu(v,\tau)=(-1)^{|v|-|\tau|}$ where $\tau$ is a subset of $v$, $|\tau|$ is the cardinality of the subset.

#### 1a. Möbius Inversion

Consider a set of $n$ variables or attributes, $v = \{X_1, X_2, X_3... X_n\}$ and adopt the sign convention of [2] to define $g$, the dual of $f$ for the set of variables, equal to the interaction information if $f$ were the entropy function, $H$.

$$g(\eta)=\sum_{\tau\subseteq\eta}\mu(v,\tau)f(\tau)=\sum_{\tau\subseteq\eta}(-1)^{|v|-|\tau|}f(\tau)\,;\,\eta,\tau\subseteq v$$

$$(2a)$$

It can easily be shown that the symmetric relation holds,

$$f(\eta)=\sum_{\tau\subseteq\eta}(-1)^{|v|-|\tau|}g(\tau)\,;\,\eta,\tau\subseteq v$$

$$(2b)$$

The relations defined in equation 2(a,b) represent a symmetric form of Möbius inversion, and the functions $f$ and $g$ can be called Möbius duals.

A chain on a lattice between elements is the set of elements on a path such that each element is greater than the adjoining element upwards and less than the adjoining element downward (including the limiting elements). The Möbius inversion is a convolution of the Möbius function with any function defined on the lattice over all its elements (subsets) between the argument subset, $\tau$, of the function and the empty set. This means all the elements, on all paths, between $\tau$ and the empty set (counting the elements only once). This range is sometimes called <u>down-set</u> of a lattice ordered by inclusion. This range is also called an ideal of the lattice. The empty set, at the limit of the range of the convolution, can be considered as the "reference element". We use the idea of a reference element in section 4 in generalizing the inversion relations.

To illustrate the realtions concretely the nodes and the Möbius function are shown graphically for three variables or elements in figure 1. When the functions in equation 2 are mapped onto the lattice for three variables, these equations represent the convolution of the lattice functions and the Möbius function over the lattice.

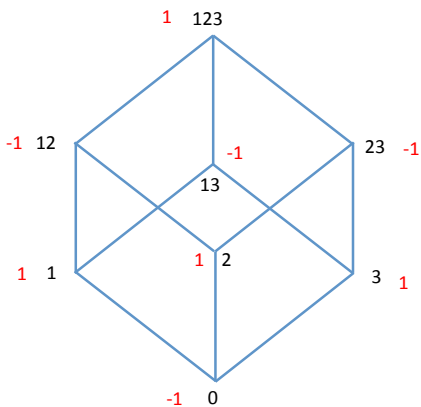

Figure 1. The Hasse diagram of the subset lattice for three variables. The numbers in black are the variable subsets, while the Möbius function $\mu(\nu, \tau)$ on this lattice (1 or -1) is indicated in red.

1b. Möbius Operators:

The convolutions with the Möbius function over the lattice in equation 2 define mappings that can be expressed as operators. The operators map one function on the lattice into another. A function on the lattice is a map of the subsets of variables at each node into the real numbers. For a set of variables, *v*, ordered by inclusion on a subset lattice, we define the Möbius *down-set operator*, ***m*** , that operates on a function on this lattice. The down-set operator is defined as an operator form of the convolution with the Möbius function: the sum over the lattice of <u>subsets of $\tau$,</u> of product of the values of the function times the Möbius function. The lower limit of this convolution is the empty set.

$$\hat{m}(f(\tau)) \equiv \sum_{\eta \subseteq \tau} (-1)^{|\eta|-1} f(\eta) = g(\tau), \tau \subseteq \nu \ , \qquad (3a)$$

Likewise, we define a Mobius *up-set operator*, $\boldsymbol{M}$, for which the sum is over the lattice of <u>supersets of $\tau$.</u> The upper limit of this convolution is the complete set.

$$\hat{M}(f(\tau)) \equiv \sum_{\eta \supseteq \tau} (-1)^{|\eta|+1} f(\eta) = h(\tau) \quad , \ \eta, \ \tau \subseteq \nu \qquad (3b)$$

Given a function, $f$, these equations define the functions $g$ and $h$, respectively: the *down-set* and *up-set* inverses or duals of $f$. The sum in the expression of eqn. 3a is the same as the symmetric form of the Möbius inversion [2]: $f$ and $g$ in eqn. 3a are interchangable, dual with respect to the down set operator (see eqn. 2a and 2b). We call the limiting subsets <u>reference subsets</u> of the operators. The up-set operator is thereby referenced to the full set, the down-set operator to the empty set.

From eqn. 3a Möbius inversion implies that applying the down-set operator twice yields the identity, $\boldsymbol{m}^2 = I$ . This is an expression of the duality. It is simple to show that the same also applies to eqn. 3b, so that $\boldsymbol{M}^2 = I$ . This idempotent property of the Möbius operators is equivalent to the symmetry in equation 2: the exchangability in these equations, or duality of the functions is exactly the same property as the idempotecy of the operators. The relationships between pairs of the dual functions, generated by the operators are shown in the diagram in figure 2. The range of the convolution operator is clear here, but this is not always true, and where it is ambiguous we use a subscript on the operator to identify the reference set. We will need this in section 4.

$$h(\tau) \xleftrightarrow{\hat{M}} f(\tau) \xleftrightarrow{\hat{m}} g(\tau)$$

Figure 2. The Möbius operators define the duality relationships between the functions on the subset lattice.

To advance this formalism further we need to define another operator on the lattice. The inversion, or complementation, operator has the effect of mapping function values of all nodes (subsets) of the lattice into the function values of the nodes that are their set complements. Viewed as a geometric space, as shown in figure 1 for 3 dimensions, the complementation corresponds to an inversion of the lattice, all coordinates mapping into their negatives through the origin at the center of the cube (for example, node 1 maps into node 23 in figure 1.) We define the operator $X$, acting on functions of subsets $\tau$ of the set $v$

$$\hat{X} f(\tau) = (-1)^{|v|} f(\tilde{\tau}): \ \tau \subseteq v, \ \tau \cap \tilde{\tau} = \varnothing, \ \tau \cup \tilde{\tau} = v$$

$$(4)$$

The sign change factor is added since inversion of the lattice also has the effect of shifting the Möbius function by a sign for odd numbers of total variables on the lattice. All operator relations on functions of $\tau$ defined so far are positive in sign. The pairwise relations among the functions and the operators shown in equation 5 then follow. The 3 and 4 variable case for equation 5 can easily be confirmed by direct calculation, and the general case is also easy to prove. The proofs are direct and follow from the Möbius inversion sums, by keeping track of the effects of each of the inversion and convolution operators.

$$h(\tau) \xrightarrow{\hat{M}} f(\tau) \xrightarrow{\hat{m}} g(\tau)$$

$$f(\tau) \xrightarrow{\hat{m}} g(\tau) \xrightarrow[\hat{X}\hat{m}]{\hat{X}\hat{M}} h(\tau)$$

$$(5)$$

These relationships among the functions determined by the mappings of the operators can be represented in a single diagram of the mappings among the three functions, as shown in figure 3, where we define the composite operators, **P** and **R**.

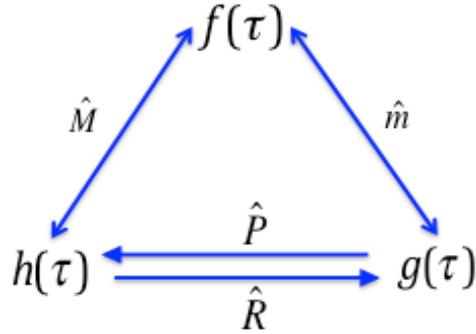

Figure 3.  Diagram of the mappings of the functions on the subset lattice into one another by the operators.  The operators **P** and **R** are:  $\hat{P} = \hat{X}\hat{M}$, $\hat{R} = \hat{X}\hat{m}$.

If we collect the operators shown in eqns. 5, add the identity operator, and calculate the full product table of the set of operators, $\{\hat{I}, \hat{m}, \hat{M}, \hat{X}\}$ and their products we determine their full algebraic properties.

The full product table is shown in Table 1.  We now ask whether this set of relations satisfies the properties of a group: closure, identity, element inverses and associativity.

| right / left | I | m | X | M | P | R |
|---|---|---|---|---|---|---|
| **I** | I | m | X | M | P | R |
| **m** | m | I | P | R | X | M |
| **X** | X | R | I | P | M | m |
| **M** | M | P | R | I | m | X |
| **P** | P | M | m | X | R | I |
| **R** | R | X | M | m | I | P |

Table 1.  The product table for the 6 operators above.  The operators P and R are defined as $\hat{P} = \hat{X}\hat{M}$, $\hat{R} = \hat{X}\hat{m}$.  The convention is that the top row is on the right and the left column on the left in the products indicated; e.g. $\hat{M}\hat{X} = \hat{R}$, $\hat{X}\hat{M} = \hat{P}$.

It is immediately clear that the product set is closed, all elements have an inverse, and the table demonstrates associativity. Thus, this set of operators indeed forms a group. Furthermore, examination of the table shows that it is isomorphic to the symmetric group $S_3$.

It is useful to show concrete representations of such relations, and we indicate in Table 2 the 3x3 matrix representation of the group $S_3$, with the one line notation of the operator effect, and the correspondance between the Möbius operators and the $S_3$ representation.

| One line Notation: (Image of String) | Matrix Representation (left action convention) | Mobius Operator |
|---|---|---|
| 123 | $\begin{pmatrix} 1 & 0 & 0 \\ 0 & 1 & 0 \\ 0 & 0 & 1 \end{pmatrix}$ | I |
| 213 | $\begin{pmatrix} 0 & 1 & 0 \\ 1 & 0 & 0 \\ 0 & 0 & 1 \end{pmatrix}$ | M |
| 132 | $\begin{pmatrix} 1 & 0 & 0 \\ 0 & 0 & 1 \\ 0 & 1 & 0 \end{pmatrix}$ | M |
| 321 | $\begin{pmatrix} 0 & 0 & 1 \\ 0 & 1 & 0 \\ 1 & 0 & 0 \end{pmatrix}$ | X |
| 231 | $\begin{pmatrix} 0 & 1 & 0 \\ 0 & 0 & 1 \\ 1 & 0 & 0 \end{pmatrix}$ | P |
| 312 | $\begin{pmatrix} 0 & 0 & 1 \\ 1 & 0 & 0 \\ 0 & 1 & 0 \end{pmatrix}$ | R |

Table 2. The 3X3 matrix representation of symmetric group $S_3$ and the corresponding Möbius operators. The one-line notation shows the permutation effects on the left.

Note that while the operators themselves depend on the number of variables, since they define convolutions, their relationships do not, so the group structure is independent of the number of variables in the lattice. For any number of variables the structure is the simple permutation group, $S_3$.

## 2. Connections to the Deltas:

The symmetric deltas were defined as overall measures of dependence using the above definitions. It is useful to illustrate the three variable case to see clearly the connections with our previously proposed information measures used to measure dependence, called deltas [4]. We defined the deltas as "differential interaction information", as in equation 1:

$$\Delta(v_{m-1}; X_m) \equiv I(v_m) - I(v_{m-1}) = -I(v_{m-1} \mid X_m) \qquad (6a)$$

The notation reflects the asymmetry of the deltas under variable permutation. It makes a difference which variable, $X_m$, is not in $v_{m-1}$. If the marginal entropies are identified with the function $f$ in equation 2, and the interaction informations with $g$, then the differential interaction information is identified with $h$. For the three variable case these examples are shown using simplifed notation,

$$h(1) = \Delta(23;1); \quad h(2) = \Delta(13;2); \quad h(3) = \Delta(12;3) \qquad (6b)$$

These three variable deltas are conditional interaction informations (within a sign), conditional mutual information in this case of three dimensions. They represent both dependence and complexity measures. This reflects a general relation, valid for any number of variables, as can easily be shown. Simplifying the notation we can express this relation using the Möbius operator as

$$\Delta(\tau; X) = \hat{M} H(X) = -I(\tau \mid X) \qquad (6c)$$

The full subset on the lattice is $\tau \cup \{X\}$ and the variable $X$ is singled out as in equations 1 and 6a. Furthermore, the convolution takes place over the set $\tau \cup \{X\}$.

Equation 6c, interpreted properly, provides a simple connection between the deltas and our operator algebra, expressing a key relation.   In terms of lattice properties, it says that:

**The Möbius *up-set* operator acting on the *join-irreducible elements* of the lattice of marginal entropies generates the conditional interaction informations, the deltas, for the full set of variables of the lattice.**

*Join-irreducible* lattice elements are all those that cannot be expressed as the join, or union, of other elements.  In this case they are the single variables.  Since the deltas are differentials of the interaction information at the top of the lattice (the argument of the function is the full set), their expression in such form is interesting.  Figure 4 illustrates the specific connection between the join-irreducible elements and deltas for the 4 variable lattice.

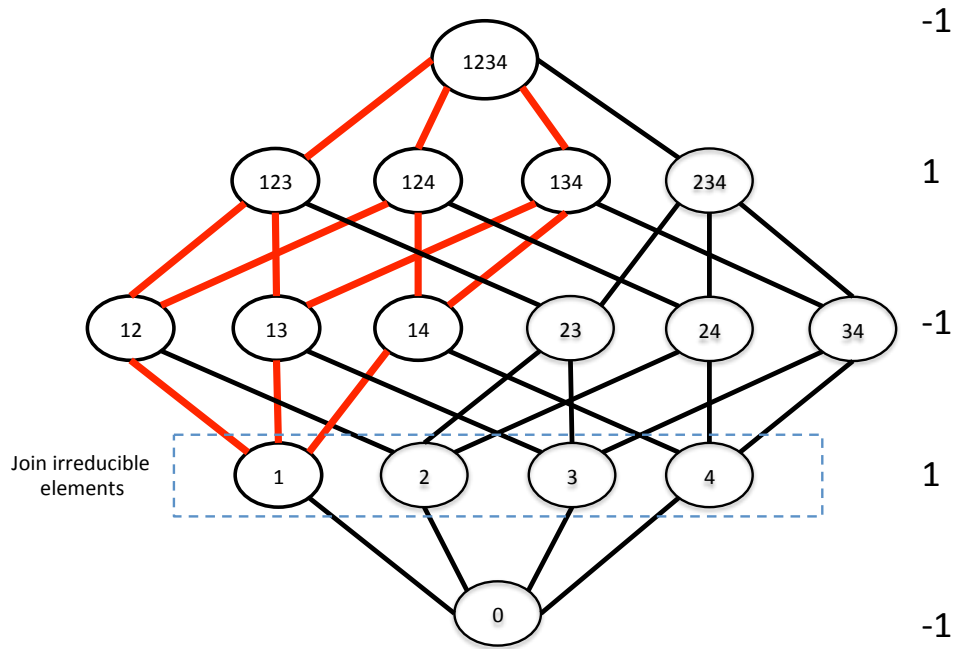

Figure 4.  The four variable lattice showing the 4 join-irreducible elements that generate the symmetric deltas as in equation 6c.  Möbius function values are shown on the right, and the red lines connect the elements of the delta function, $\Delta(234;1)$, which form a 3-cube.

A general statement of this connection is that the differential of one function on the lattice corresponds to the up-set operator on another function of the join-irreducible elements.    This is a general property of the algebraic structure of the subset lattice.  Written in terms of the functions related by the inversions, and using the same set notation as above, $X$ indicating a join-irreducible element, we can state this general result as follows.

If   $g(\tau) = \hat{m}f(\tau)$  and $X$ is a join-irreducible element of lattice, then

$$\hat{M}f(X) = h(\tau; X) = g(\tau \,|\, X)$$

(7)

where the final term is a conditional form of the $g$ function in which $X$ is instantiated.

We have previously proposed the symmetric delta (the product of all variable permutations of the delta function, $h$) as a measure of complexity, and of collective variable dependence [6].    The symmetric delta expression, created by taking the product of the individual deltas is seen to be the product of the results of the up-set operator acting on the functions of *all of the join-irreducible elements of the entropy lattice*.   Note also that by equation 1 both the conditional entropies and conditional interaction informations, since they correspond to the differentials, imply a path independent chain rule.  Note that the product in this case is over all the join-irreducible elements only.

$$\overline{\Delta}(\tau) = -\prod_{X \in \tau} \hat{M}H(X)$$

(8)

Note that these kinds of differential functions include more than just those keyed on the join-irreducible elements.  We have used only the deltas, measures of dependency, but the general function set includes a number of others.

### 3.  Symmetries reveal a wide range of new relations

There are a number of other relations implied by this system of functions and operators.  Examination of equation 1 and comparision with 6c shows that delta is also related to what we can call the differential entropy.   We define this quantity as $\delta H(v_n) \equiv H(v_n) - H(v_{n-1})$  the change in the entropy of a system when we consider an additional variable.   Applying the down-set operator to this quantity, we obtain

$$\hat{m}\delta H(v_n) = \hat{m}(H(v_n) - H(v_{n-1})) = I(v_n) - I(v_{n-1}) = -I(v_n \,|\, X_n)$$
$$\hat{m}\delta H(v_n) = \hat{M}(H(X_n)) \tag{9}$$

where $X_n$ is the element that is the difference between the sets $v_n$ and $v_{n-1}$.  We can consider $\delta$ as an operator, but note that it does not define a convolution over elements of the lattice.  If we cast $\delta$ as an operator then we note that $\boldsymbol{\delta}$ and $\boldsymbol{m}$ commute.  The duality between $H$ and $I$ implies a dual version of equation 9 as well. If we apply other operators to the expression in equaiton 9 we find another set of relations among these marginal entropy functions.  For example, this remarkable symmetry emerges.

$$\delta H(v_n) = \hat{m}\hat{M}H(X_n) = \hat{X}\hat{m}H(X_n) = \hat{R}H(X_n)$$
$$H(X_n) = \hat{P}\delta H(v_n) \tag{10}$$

These equations relate functions over the lattice to functions of join irreducible elements.

There are further symmetries in this set of  information functions.   Consider the mapping diagram of figure 3.  If we define a function which is simply the delta function with each node mapped into its set complement, that is, acted on by the lattice inversion operator, we have

$$\Phi \equiv \hat{X}\hat{\Delta}; \quad \hat{X}\hat{m}\Phi = \hat{X}\hat{m}\hat{X}\Delta = H \tag{11}$$

Then these functions occupy different positions in the mapping diagram as seen in figure 5.  Several other such modifications can be generated by similar operations.

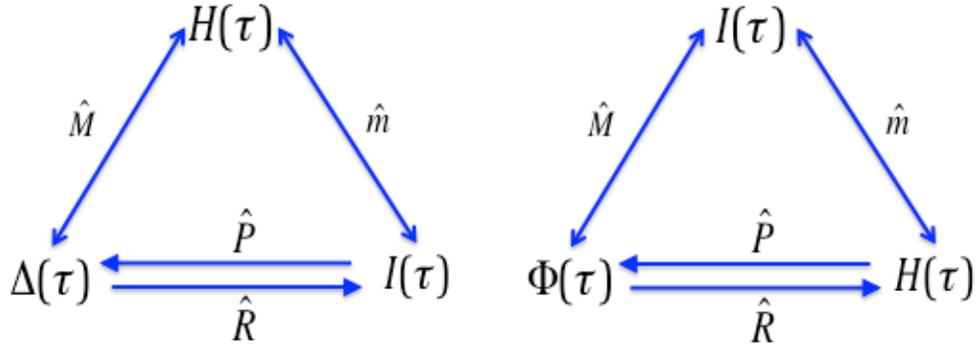

Figure 5. A simple modifcation of one of the functions by lattice inversion modifies the postion of functions in the mapping diagram. The original diagram is on the left, the result of $\Delta$ modified by inversion is on the right. Note that the idempotent property of **M** is applied to the $\Phi$ function relation.

There are a large number of similar relations that can be generated by such considerations.

There are other information-based measures that we can express using the operator algebra. Because it is a widely used measure for multi-variable dependence we will now examine the example of multi-information, also called "total correlation". Multi-information, $\Omega$, is defined as the difference between the sum of the entropies of each of $n$ variables from the set $v_n = \{ X_i \}$ and the joint entropy of the set , $H(v_n)$ [10]. It is often used because it is always postive and goes to zero when all the variables are independent. It is a kind of conglomerate of dependencies among members of the set $v_n$ .

$$\Omega(v_n) \equiv \sum_{X_i} H(X_i) - H(v_n) \qquad (12a)$$

In terms of entropy functions on the lattice elements, $\Omega$ as expressed in this equation can be thought of as the sum of the join-irreducible elements, minus the "top element" or join of the entropy lattice. To apply the down-set operator to the terms in the equation we must remember that both terms on the right are functions of $v_n$ on the lattice, even though the first term is a sum over the members of this set. If we calculate the convolution over the **$\Omega$** function, we have

$$\hat{m}\Omega(\upsilon_n) = -I(\upsilon_n) \qquad (12b)$$

The multi-information of a single variable is 0 (realizing that $H_i - H_i$). This causes a problem if we apply the down-set operator to both sides of (12b), however, as the sum of the single entropies is lost. This is simply the result of the definition of this function. It is a compound function in the sense that it is not a simple convolution, including functions at the limit of the range of the convolution, and is therefore in a different class than those functions of only the subset elements. They behave aberrantly under the Mobius operators. The application of the up-set operator to the multi-information function on the lattice gives us

$$\hat{M}\Omega(X_i) = -\Delta(\upsilon_{n-1}; X_i)$$

$$(12c)$$

In spite of this inconsistency with the definiton of the function, the set of relations in equation 12 can elucidate some of the properties of the multi-information. Formally these kinds of compound functions must, however, be excluded.

A closer look at the multi-information function: if the full set is $\upsilon_n$, how does the function on other lattice elements behave under these operators? We defer a general treatment for later, but illustrate here the connection for the 3-D case. Consider the $\Omega$ function over the 3-cube, and note that the values of the function for all join-irreducible elements is zero. If we calculate $\hat{M}\Omega(X_3)$ then we have

$$\hat{M}\Omega(X_3) = -\Omega(X_2, X_3) - \Omega(X_1, X_3) + \Omega(X_1, X_2, X_3) \qquad (13)$$

The first two terms on the right are the mutual informations for the indicated variables, and the righthand side is indeed $\Delta(X_1, X_2; X_3)$. Written in another way we have: $\Delta(X_1 X_2; X_3) = I(X_1 X_3) + I(X_2 X_3) - \Omega(X_1 X_2 X_3)$.

## 4. Relation to probability densities

4a. Conditional log Likelihoods and Deltas

Note that the differential entropy (equations 9 and 10) is the same as conditional entropy by the chain rule. Writing this in terms of the probability distributions, using the definitions of the joint entropies and the probability chain rule, gives

$$\delta H(v_n) \equiv H(v_n) - H(v_{n-1})$$
$$\delta H(v_n) = -\left\langle \ln \frac{P(v_n)}{P(v_{n-1})} \right\rangle = -\left\langle \ln P(X_n \,|\, v_{n-1}) \right\rangle = H(X_n \,|\, v_{n-1}) \qquad (14a)$$

For simplicity of notation we define $\pi$ as this expectation value. We have

$$\pi(X_n \,|\, v_{n-1}) \equiv -\left\langle \ln \frac{P(v_n)}{P(v_{n-1})} \right\rangle = \left\langle \ln P(v_{n-1}) - \ln P(v_n) \right\rangle \qquad (14b)$$

From 14 we see that $\pi$ is a conditional log likelihood function, which is the same as the conditional entropy, and the difference between two entropies. These relations are the consequence of the definition of entropy and the probability chain rule. By applying the down-set operator, ***m***, to this quantity we generate some interesting relations. As seen in equations 9 and 10, the result of this operation is the delta, the conditional interaction information,

$$\hat{m}\pi(X_n \,|\, v_{n-1}) = \hat{m}\delta H(v_n) = \hat{M}H(X_n) = -I(v_{n-1} \,|\, X_n) = \Delta(v_{n-1}; X_n)$$
$$(15)$$

Expressing this in another way, using the group table, we have the expressions from equation 10, $\delta H(v_n) = \hat{m}\hat{M}H(X_n) = \hat{X}\hat{m}H(X_n) = \hat{R}H(X_n)$ and therefore

$$\pi(X_n \,|\, v_{n-1}) = -\left\langle \ln P(X_n \,|\, v_{n-1}) \right\rangle = \delta H(v_n) = \hat{R}H(X_n) \qquad (16a)$$

The expected value of the log of the probability of a given, single variable, conditioned on the other variables in the subset, can therefore be expressed simply in terms of Möbius operators acting on the entropy functions of a lattice. The major result of this section, expressed as the symmetric delta is

$$\overline{\Delta}(v_n) = \prod_{all\,choices\,of\,X_n} \Delta(v_{n-1};X_n) = \prod_{all\,choices\,of\,X_n} \hat{m}\pi(X_n|v_{n-1}) \qquad (16b)$$

The relation of the $\pi$'s to the deltas is clear. The subsets of the variables under consideration then can generate a series of conditional log likelihoods (CLL's) for $|v_m|=m$, $\{\pi(X_m|v_{m-1})\}$ for $m \geq 2$. The Bayesian approximation for dependencies among variables is realized in the case $m = 2$, where all CLL's are approximated by those with single conditonal variables. In this case (using simplified notation)

$$\pi(2|1) = H_{12} - H_1$$
$$\pi(3|1) = H_{13} - H_1 \qquad (17a)$$

and we have for the three variable case

$$\Delta(23;1) = H_1 - H_{12} - H_{13} + H_{123} = -\pi(2|1) + \pi(2|13)$$
$$\Delta(23;1) = -\pi(3|1) + \pi(3|12) \qquad (17b)$$

There are two different ways to express deltas as sums of the $\pi$'s. Several things follow from these considerations. First, there is a correspondence, albeit expressible in several ways, between the deltas and CLL's. Second, since the group table for the Möbius operators exhibits several different, equivalent operators, $\hat{R} = \hat{m}\hat{M} = \hat{X}\hat{m} = \hat{M}\hat{X} = \hat{P}^2$, we can express the correspondence between $\Delta$ and the CLL's in several equivalent ways. These expressions provide direct links with other information functions.

A possible approach for finding relations predictive of a variable from the information in a data set is suggested by the above considerations. The general problem can be defined as how to determine the "best" prediction formula for the value of one variable in the set, say $X_1$, from analysis of a data set of all variables. We sketch the suggested approach here. Step one in a high level description of the process, is to define the maximum degree of dependence to consider (the number of variables involved.) Step two is to use the symmetric deltas to determine the set of variables that are dependent on one another [9]. Step three is to find the maximum expected CLL, from the set $\left\{\pi(X_1|X_i), \pi(X_1|X_iX_j), \pi(X_1|X_iX_jX_k)...\right\}$ by calculating the expectations of the entropy differentials. Note that the specifc, expected entropy differences tend to zero as the dependence of the single variable, $X_1$ , on the others increases. Finally, once the "best" likelihood function is found, a predictive function is estimated based on the data can be made: an estimate of the probabilities of $X_1$ conditioned on all the other members of the set. Practical ways of calculating the predictive functions, and algorithms for doing so, is similar to well studied problems in Machine Learning and statistics and will not be considered here. The general framework for inference is nonetheless clear. This procedure is reminisent of the Chow-Liu algorithm [12] which is entirely pairwise. This approach can provide a method for generating predictive rules from large, multivariable data sets. We will develop this appraoch further in another article.

## 5. Generalizing the Möbius operators

The up-set and down-set operators, **M** and **m**, generate convolutions over paths from each element or subset to the "top" (full set) or to the "bottom" (empty set) respectively. The convolutions are therefore either "down", towards <u>included</u> subset elements, or "up" toward <u>including</u> subsets. The paths over which the sums of the product of function and Möbius function are taken to form the convolution are clear and are defined by the subset lattice for these two operators. No element is included more than once in the sum. Moreover, the sign of the Möbius function is the same across all elements at the same distance from the extreme elements.

We can generalize the Möbius operators by defining the range of the convolution, the end elements of the paths, to be <u>any</u> pair of elements of the lattice. Two elements are required the starting, or subject element, and an ending element. We call such an ending element a reference element and associate it with an operator. Instead of the up-set operator, with the full set $v$ as its reference element, we could designate an arbitrary subset element like {1,2} as the reference and thereby define another operator. Consider a lattice of the full set $v$, where $\eta$ designates a reference element. We then define the corresponding operator $F_\eta$, acting on a function, $f(\tau)$, as

$$F_\eta f(\tau) = \sum_{\substack{\varsigma \text{ on all shortest paths} \\ \text{between } \tau \text{ and } \eta}} (-1)^{|\nu|-|\varsigma|} f(\varsigma) \qquad (18)$$

In this lattice there are multiple shortest paths between any two elements, since the subset lattice is simply a hypercube. Like the upset and down set operators we only include each element once. The two extreme reference elements, the empty set and the full set, yield the down-set and up-set operators respectively

$$F_0 f = \hat{m} f$$
$$F_\nu f = \hat{M} f \qquad (19)$$

The reference element establishes a relation between the lattice sums and the Möbius function. It is the juxtaposition of the lattice, anchored at the reference element, to the Möbius function that defines the symmetries of the generalized algebra. Note that we now have the possibility of including elemeents that are not ordered along the paths by inclusion. The convolution between {1} and {2,3} for the 3-cube lattice shows this clearly as it inclues {1,2}, {2} and the empty set.

The products of operators can easy be calculated for the 3 and 4 element sets. We can identify some similarities of these general operators to the operators **M** and **m**. First, we note that the operators $F_\mu$ are all idempotent. This is easy to calculate for the 3D and 4D case, and to derive using the relations indicated in equations 18 and 19. The idempotent property then means that there are pairs of functions that are related by each general Möbius operator – a generalized Möbius inversion on the inclusion lattice, a generalized duality. Furthermore, the products exhibit some other familiar symmetries. For all $\mu, \eta \subseteq \nu$ The relationship of the products involves the operator *X*, which in the geometric metaphor affects a rotation of the hypercube (subset lattice.)

$$F_\mu F_\eta = X F_\eta F_\mu X$$
$$F_\mu F_\eta = \tilde{F}_\eta \tilde{F}_\mu \qquad\qquad \text{(20a)}$$

where we define the complement operator here as $\tilde{F}_\mu \equiv \hat{X} F_\mu \hat{X}$. This expression can be shown to be equivalent to

$$F_\mu \tilde{F}_\eta = F_\eta \tilde{F}_\mu$$
$$\text{(20b)}$$

One can be derived from the other. There are a number of other symmetries including the following: if $\tilde{\mu}$ is the complement of $\mu$ then

$$F_\mu = -\tilde{F}_{\tilde{\mu}} \qquad\qquad \text{(21)}$$

There are also symmetries among these operators that involve combinations of operators and lattice functions. A notable relationship that involves a subset and its complement is the following

$$F_\mu f(\tilde{\mu}) = F_\eta f(\tilde{\eta}) \qquad\qquad \text{(22)}$$

which is true for any subsets $\mu$ and $\eta$ and their complements. This expression is seen to describe the convolution over <u>all</u> subsets of the entire lattice, so the proof is trivial.

The full group structure of the general operator algebra is more complex than the group defined by the up-set and down-set operators as there are many more operators, defined by the full range of reference elements. Secondly, the Möbius function is generally not aligned with the reference elements as it is for the up-set and down-set operators, so this symmetry is broken. Remarkably, the symmetry of the subgroups determined by pairs of complementary subsets are nonetheless preserved, remaining isomorphic to $S_3$ (seen to be true for the 3D, and 4D case by direct calculation, and it appears to be generally true, though we do not yet have a proof.) The relations between these pairs of functions on the lattice is described by the diagram in figure 6. It appears that the sets of three functions, specific to a reference set $\eta$, with the operators that map one into the other exhibit the same overall symmetries reflected in the group $S_3$. The pairs of operators identified with a subset and its complement are the key elements of the group. This is because this particular combination of operator and function defines a convolution over the entire set, $v$. This identity therefore includes the specific up-set and down set relations, and is equal to the interaction information if $f$ is the entropy function.

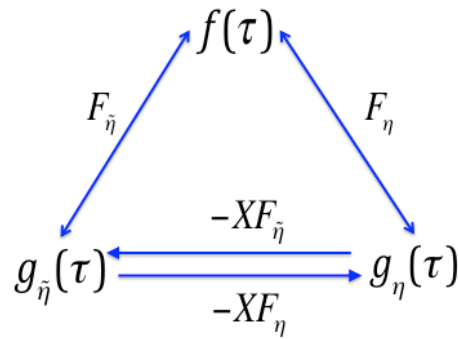

Figure 6. Generalized Möbius operator relations. A diagram of the relations among the functions as determined by the operators. The upper two arrows represent the generalized Möbius inversion relations. The $S_3$ structure is reflected in the similarity with the diagram of figure 3. Note that when $\eta=\varnothing$        figure becomes identical to figure 3.

We ask now if sums of such operator-function pairs can decompose the full convolution. This decomposition can be addressed by asking this specific question: are there sums of operators on functions that add up to specific convolutions of one operator on one function, and if so what are they? The decomposition of the hypercube into sub-lattices can be shown to be equivalent to the process of finding these decompositions. We will not deal with the general decomposition relations here but show them for {1,2,3} and {1,2,3,4}. First, an example. The following decomposition results from decomposing the 3-cube Hasse diagram into two squares (2D hypercubes), which is done by passing a plane through the center of the cube in one of three possible ways

$$F_0\, f_{123} = F_2\, f_{123} + F_0\, f_{13} \qquad (23)$$

Each of the two terms on the right could be expressed as operator terms in four ways (each of the four subsets of the four nodes.)  Each of the four leads to the same set of functions, but it is a distinct operator expression.  There are three ways of decomposing the cube into two squares, so there are a total of 48 decompositions of the full 3-set convolution.  For the 4-set decomposition, there are three ways of decomposing the 4-hypercube into 2 cubes, so the total number of possible decompositions is  3 x 48 x 48 = 6912.    The general expression for the number of possible such decompositions for a set $\tau$, where $|\tau| = n$  appears to be    $3^{(2^{n-2}-1)}4^{2^{n-2}}$ .

## 6. Discussion

Many diverse measures have been used in descriptions of order in complex systems and as data analysis tools [1-9].  While the mathematical properties and relationships among these information-related measures are of significant interest in several fields, there has been, to our knowledge, no systematic examination of the full range of relationships and no unification of this diverse range of functions into a single formalism as we do here.  Beginning with the duality relationships, based on Möbius inversions of functions on lattices, we define a set of operators on functions on subset inclusion lattices that map the functions into one another.   We show here that they form a rather simple group, isomorphic to the symmetric group $S_3$.  A wide range of relationships among the set of functions on the lattice can be expressed simply in terms of this operator algebra formalism.   When applied to the information-related measures they can express a wide range of relationships among various measures, providing a unified picture and allowing new ways to calculate one from the other using the subset lattice functions.  Much is left to explore in the full range of implications of this system, including algorithms for prediction, and other practical matters for dealing with complex data sets.

We are able to establish points of connection with other areas where lattices are useful. Since any distributive lattice is isomorphic to the lattice of sets ordered by inclusion, all the results presented here apply to any system of functions defined on a distributive lattice [13], so this unification extends well beyond the information measure functions[1].

The relationships shown here unify, clarify, and can serve to guide the use of a range of measures in the development of the theoretical characterization of complexity, and in the algorithms and estimation methods needed for the computational analysis of multi-variable data. We have addressed the relationships between the interaction information, the deltas (conditional interaction information), and the underlying probability densities. We find that the deltas can be expressed as Möbius sums of conditional entropies, the multi-information is simply related by the operators to other information functions, and we made an initial connection to the maximum entropy method as well.

We also note that Knuth has proposed generalizations of the zeta and Möbius functions that define *degrees of inclusion* on the lattices [11]. Knuth's formalism, taken with ours, would lead to a more general set of relations, and add another layer of complexity, uncertainty or variance to the information-related measures. This could be particularly useful in developing future methods for complexity descriptions and data analysis. From the simple symmetries of these functions and operators it is clear there is more to uncover in this complex of relationships. The information theory-based measures have a surprising richness and internal relatedness in addition to their practical value in data analysis. The full range of possible relationship in applications that have used Möbius pairs of functions remains to be explored. Since the information-related functions have been directly linked to interpretations in algebraic topology [13] it will also be interesting to explore the topological interpretation of the Möbius operators.

---

[1] This is according to the theorems of Stone and Priestly. Distributive lattices are widespread and include the following: every Boolean algebra is a distributive lattice; the Lindebaum algebra of most logics that support conjunction and disjunction is a distributive lattice; every Heyting algebra is a distributive lattice, every totally ordered set is a distributive lattice with *max* as join and *min* as meet. The natural numbers also form a distributive lattice with the greatest common divisor as meet and the least common multiple as join (this infinite lattice, however, requires some extension of the equivalence proof.)

**Acknowledgements:** We gratefully acknowledge support by the National Science Foundation, EAGER grant (IIS-1340619), by the Pacific Northwest Research Institute, and the Bill and Melinda Gates Foundation. In the early stages of this work the Luxembourg Centre for Systems Biomedicine (LCSB) of the University of Luxembourg, also provided significant support.